\documentclass[preprint,prd,tightenlines,bibnotes]{revtex4-1}

\usepackage{graphicx} % Include figure files
\usepackage{dcolumn}  % Align table columns on decimal point
\usepackage{colordvi}
\usepackage{color}
\usepackage{epstopdf}
\usepackage{amssymb}
\usepackage{amsmath}
\usepackage{url}
\graphicspath{{ps}}
\usepackage{hyperref}
\usepackage{tabularx}
\usepackage{placeins}
\usepackage{adjustbox}

\input belle2sym.tex

\begin{document}

%place for definitions and newcommands
\def\belletwo {\it {Belle II}}

\vspace*{-3\baselineskip}
\resizebox{!}{3cm}{\includegraphics{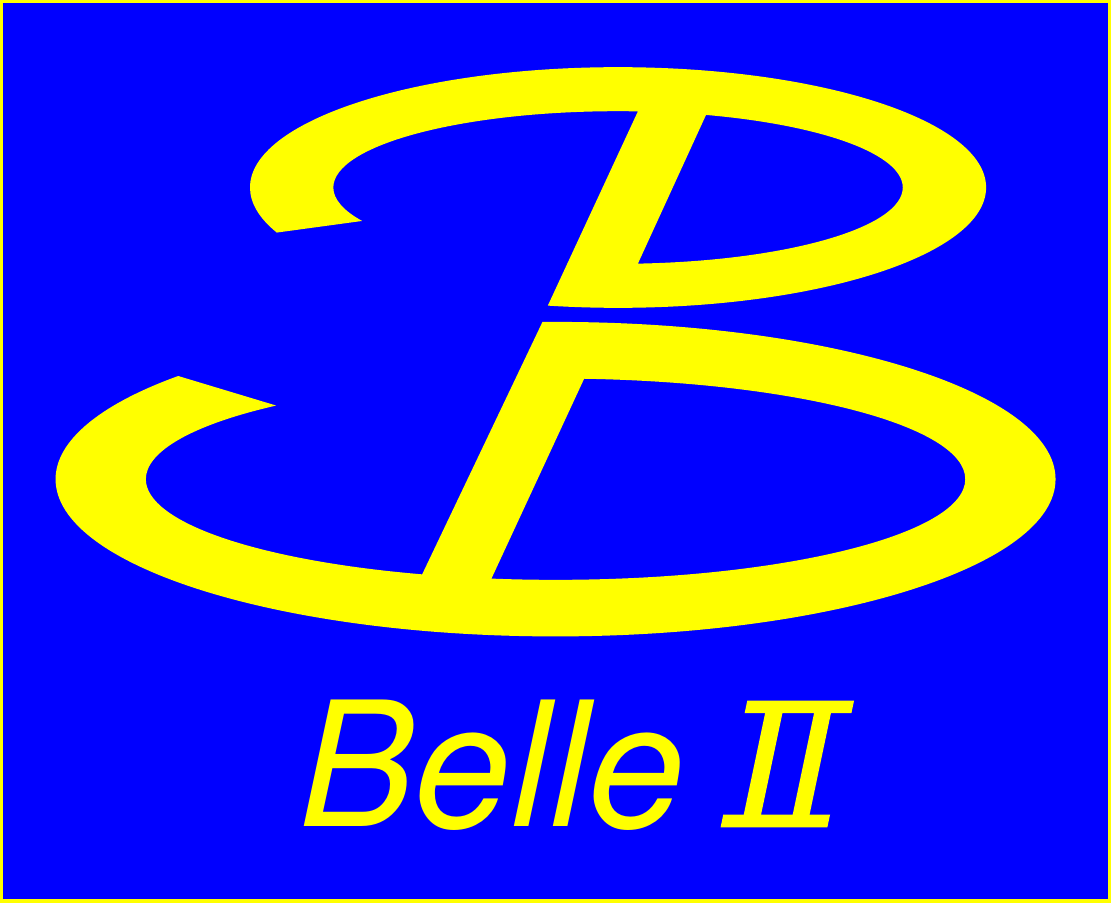}}

\vspace*{-5\baselineskip}
\begin{flushright}
BELLE2-CONF-PH-2022-008\\
September 19, 2022
\end{flushright}

\title { \quad\\[0.5cm] Measurement of the branching fractions and {\it CP} asymmetries of $\Bp \to \pip \piz$ and $\Bp \to \Kp \piz$ decays in 2019-2021 Belle II data}

%%% Paper:    (2022 conference papers)
%%% Journal:  (2022 conferences)
%%% Updates:
%%% June 21, 2022 - plain format
%%% ====================================================================
%%% Use \input{authors-conf2022-plain} to insert this material into your latex file.
\collaboration{The Belle II Collaboration}
  \author{F. Abudin{\'e}n}
  \author{I. Adachi}
  \author{K. Adamczyk}
  \author{L. Aggarwal}
  \author{P. Ahlburg}
  \author{H. Ahmed}
  \author{J. K. Ahn}
  \author{H. Aihara}
  \author{N. Akopov}
  \author{A. Aloisio}
% ...

\begin{abstract}
We determine the branching fractions ${\mathcal{B}}$ and {\it CP} asymmetries ${\mathcal{A}_{{\it CP}}}$ of the decays ${\Bp \to \pip \piz}$ and ${\Bp \to \Kp \piz}$.
The results are based on a data set containing 198~million bottom-antibottom meson pairs corresponding to an integrated luminosity of $190\invfb$ recorded by the Belle~II detector in energy-asymmetric electron-positron~collisions at the $\FourS$ resonance. 
We measure 
\begin{center}
${\mathcal{B}(\Bp \to \pip \piz) = (6.12 \pm 0.53 \pm 0.53)\times 10^{-6}}$,

${\mathcal{B}(\Bp \to \Kp \piz) = (14.30 \pm 0.69 \pm 0.79)\times 10^{-6}}$,

${\mathcal{A}_{{\it CP}}(\Bp \to \pip \piz) = -0.085 \pm 0.085 \pm 0.019}$,  

${\mathcal{A}_{{\it CP}}(\Bp \to \Kp \piz) = 0.014 \pm 0.047 \pm 0.010}$, 
\end{center}
where the first uncertainties are statistical and the second are systematic.
These results improve a previous Belle~II measurement and agree with the world averages. 
\end{abstract}


\maketitle

{\renewcommand{\thefootnote}{\fnsymbol{footnote}}}
\setcounter{footnote}{0}

\section{Introduction}

Studies of hadronic charmless $B$ decays give access to the angle $\alpha / \phi_2 \equiv \mathrm{arg}(-\frac{V_{td}V_{tb}^{*}}{V_{ud}V_{ub}^{*}})$, where $V_{ij}$ are the elements of the Cabibbo-Kobayashi-Maskawa (CKM) quark-mixing matrix. 
This is the least known angle of the unitarity triangle.  
Charmless decays also probe contributions from dynamics beyond the standard model (SM) in processes mediated by loop decay-amplitudes.

The angle $\alpha$ can be measured in neutral $B$ meson decays governed by $b \to u \overline{u} d$ transitions, such as $\Bz \to \pip \pim$, using interference between the amplitude for a direct decay and for a decay following $\Bz-\Bzb$ flavor oscillation \cite{CC}. However, the determination of $\alpha$ is straightforward only if one decay amplitude contributes.
As both tree and penguin decay-amplitudes contribute to hadronic two-body charmless $B$ decays, the measured value of $\alpha$ is shifted by a weak and strong phase. 
This phase can be removed by an analysis of multiple isospin-related $\B \to \pi \pi$ decays.
The angle $\alpha$ can be determined if the two isospin-related decays $\Bz \to \piz\piz$ and $\Bp \to \pip\piz$ are measured in addition to $\Bz \to \pip \pim$~\cite{Charles_2017}. 

In addition, isospin symmetry can be employed to build sum-rules, \emph{i.e.}, linear combinations of branching fractions and direct {\it CP} asymmetries, to test SM predictions.
For the set of $B \to K\pi$ decays, $\Bz \to \Kp\pim$, $\Bp \to K^0\pip$, $\Bp \to \Kp\piz$, and $\Bz \to K^0\piz$, there is a reliable and sensitive test based on comparing the observed value of the sum-rule and the SM expectation~\cite{PhysRevD.59.113002}.
For $\Bz \to K^0\piz$, the sum-rule predicts a direct {\it CP} asymmetry of $-0.138 \pm 0.025$ \cite{Aaij_2021}, using world averages for the other sum-rule inputs \cite{HFLAV}.
The average of the direct measurements is $-0.01 \pm 0.10$ \cite{HFLAV}.
Probing the sum-rule with higher precision will provide a stringent null test of the SM \cite{Gershon_2007}. 
While the sensitivity of the test is dominated by the $\Bz \to K^0\piz$ precision, improvements in all inputs will contribute in the longer term. 
Belle II is the only experiment capable of measuring jointly, and within a consistent experimental environment, all isospin-related decays.

Herein, we report measurements of the branching fractions and {\it CP} asymmetries of ${\Bp \to \pip\piz}$ and ${\Bp \to \Kp\piz}$ decays, which provide information on $\alpha$ and on the isospin sum-rule, respectively.
We use a data set corresponding to $190\invfb$ integrated at the $\FourS$ resonance, and $18\invfb$ integrated $60\mev$ below the resonance, with the Belle II detector in energy-asymmetric electron-positron ($\ep\en$) collisions provided by the SuperKEKB accelerator.

Reconstructed decay candidates are selected with a multivariate algorithm trained to suppress the dominant background from $\ep\en \to \qqbar$ (continuum-background), where $q$ is a $u$, $d$, $c$, or $s$ quark. 
A three-dimensional fit is used to determine signal yields. 
Simulated events are used to study the sample composition, determine signal efficiencies, and for fit modeling. 
Fit models are corrected for possible data-simulation discrepancies using ${\Bz \to \Dzb (\to \Kp \pim) \piz}$ and ${\Bp \to \Dzb (\to \Kp \pim \piz) \pip}$ control channels. 
This measurement improves upon a previous Belle~II measurement~\cite{Belle-II:2021gkm} due to a three-fold larger data set and improved analysis techniques. 
The latter include the addition of another signal-to-background discriminating variable in the fit to determine the signal yields, and improved methods to correct the simulation using control data. 

\section{The Belle II Detector and Simulation}

The Belle~II detector consists of several subsystems arranged in a cylindrical structure around the beam pipe \cite{Belle-II:2010dht}.
In the Belle II coordinate system, the $x$-axis is defined to be horizontal and to be pointing outside of the accelerator main-rings tunnel, the $y$-axis is vertically upward and the $z$-axis is defined in the direction of the electron beam.
The azimuthal angle $\phi$ and the polar angle $\theta$ are defined with respect to the $z$-axis.
Radial displacement $r$ is defined in the $x$-$y$ plane ($r = \sqrt{x^2 + y^2}$).
The nominal interaction point is the origin of the coordinate system.
The tracking system consists of a two-layer silicon-pixel detector, surrounded by a four-layer double-sided silicon-strip detector and a 56-layer central drift chamber.
For this data sample, the second pixel layer is only partially instrumented and covers one sixth of the azimuthal angle.
The combined silicon-tracking system provides an average vertex resolution along the beam 
direction of approximately $25\;\mum$ 
for fully reconstructed \Bz mesons.
A time-of-propagation counter and an aerogel ring-imaging Cherenkov counter cover the barrel and forward endcap regions of the detector, respectively, and provide charged-particle identification.
An electromagnetic calorimeter (ECL) fills the remaining volume inside a 1.5~T superconducting magnet that produces an uniform axial field. 
The ECL consists of a central barrel section and two annular endcaps.
The calorimeter covers the polar angle regions $12.4^\circ < \theta < 31.4^\circ$ (forward endcap), $32.2^\circ < \theta < 128.7^\circ$ (barrel), and $130.7^\circ < \theta < 155.1^\circ$ (backward endcap).
It measures the energy of photons and electrons and provides input for particle identification.
A dedicated system to identify $K^0_\mathrm{L}$ mesons
and muons is installed in the outermost part of the detector.

The data are processed using the Belle~II analysis software framework
\cite{Kuhr:2018lps}. 
Simulated events are generated using \texttt{KKMC} for quark-antiquark pairs from $e^+ e^-$ collisions \cite{Jadach:1999vf}, \texttt{PYTHIA8} for hadronization \cite{Sjostrand:2014zea}, \texttt{EVTGEN} for the decay of the generated hadrons \cite{Lange:2001uf}, and \texttt{GEANT4} for the detector response \cite{GEANT4:2002zbu}.
The simulation includes simulated beam-induced backgrounds~\cite{Lewis:2018ayu}.

\section{Selection and Reconstruction}

We reconstruct $\Bp \to \pip \piz$ and $\Bp \to \Kp \piz$ signal decays as well as ${\Bz \to \Dzb (\to \Kp \pim) \piz}$ and ${\Bp \to \Dzb (\to \Kp \pim \piz) \pip}$ control channels. 
Neutral pion candidates are reconstructed from pairs of photon-candidates.
Photon-candidates are detected in the ECL as sets of adjacent channels with a signal (cluster) not associated to the extrapolation of a trajectory of a charged particle (track). 
Clusters are required to satisfy an energy requirement depending on their polar angle (cluster energy greater than $ 0.12 \gev$ in the forward endcap, $ 0.03 \gev$ in the barrel and $ 0.08 \gev$ in the backward endcap).
Furthermore, the three-dimensional angular separation between the clusters is required to be less than 0.9 rad and the azimuthal separation is required to be less than 1 rad.
The magnitude of the cosine of the angle between the photon-candidate direction in the $\piz$ candidate rest frame and the $\piz$ candidate direction in the laboratory frame is required to be less than 0.98 to suppress combinatorial background from collinear soft photons.
Finally, the mass of the $\piz$ candidates must be in the range ${0.121 < m({\gamma\gamma}) < 0.142\gevcc}$. 

Charged pion and kaon candidates are reconstructed from track candidates. 
All tracks are required to have an absolute distance of closest approach to the interaction point less than 2 cm in the $z$ direction and less than 0.5 cm radially.
In addition, tracks must be within the drift chamber acceptance and have 20 or more associated measurement-points (hits). 
Reconstructed events are divided into disjoint pion- and kaon-enriched samples via a binary pion-kaon particle-identification requirement. 
In the kaon-enriched sample, 80\% of $\Kp \piz$ candidates are correctly identified, while 20\% of them are misidentified. 
For $\pip \piz$ candidates, 88\% are correctly identified in the pion-enriched sample and the remainder are misidentified.
For the control modes, the mass of the neutral $D$ candidates must be in the ${1.84 - 1.88\gevcc}$ range.
Requirements on the beam-constrained mass $M_{\text{bc}}$ and energy difference $\Delta E$ are applied. 
The beam constrained mass is defined as $M_{\text{bc}} = \sqrt{(E^{*}_{\text{beam}})^2 - (p^{*}_{B})^2}$, where $E^{*}_{\text{beam}}$ is the beam energy and $p^{*}_{B}$ is the momentum of the reconstructed $B$ \cite{COM}.
The $B$ momentum is calculated as $p^{*}_\B = p^{*}_{h^+} + \sqrt{(E^{*}_{\text{beam}} - E^{*}_{h^+})^2 - M^2_{\piz}} \cdot \frac{p^{*}_{\piz}}{|p^{*}_{\piz}|}$,
where $p^{*}_{h^+}$ is the momentum of the charged hadron in the $B$ candidate decay, $E^{*}_{h^+}$ is its energy, $p^{*}_{\piz}$ is the $\piz$ momentum, and $M_{\piz}$ is the known $\piz$ mass \cite{modMbc}. 
The energy difference is defined as $\Delta E = E^{*}_{B} - E^{*}_{\text{beam}}$, where $E^{*}_{B}$ the energy of the $B$ candidate.
For the signal channels, $B$ candidates are retained if they satisfy $M_{\text{bc}} > 5.22\gevcc$ and $-0.3 < \Delta E < 0.3\gev$.
For the control channels, the same $M_{\text{bc}}$ requirement is applied, however, the $\Delta E$ requirement is more stringent, $-0.15 < \Delta E < 0.3\gev$.
Since the fraction of events with multiple candidates is only 1\%, we retain all candidates. 
In Sec.~\ref{sec:syst}, a systematic uncertainty is included for this choice.

\section{Continuum Suppression}
\label{subsec:bdt}

After the baseline selection, a large contribution from continuum-background remains. 
We train a discriminator to isolate the signal using a boosted decision tree (BDT). 
The variables included in the training are event shape \cite{fw, CLEO:1995rok, Belle:ksfw}, flavor tagger output \cite{Belle-II:2021zvj}, and vertex fit variables typically used at $B$ factories.
Variables having a high correlation with $\Delta E$ and $M_{\text{bc}}$ are excluded from training to avoid correlations between the BDT output and $\Delta E$ or $M_{\text{bc}}$, since these three variables are used to determine the signal yield. 
Separate discriminators are trained and tested for the two signal channels using simulated events.
A total of 36 variables are used as input to each BDT.
We identify the requirement on the BDT output that maximizes $\text{S}/\sqrt{\text{S} + \text{B}}$, where $\text{S}$ and $\text{B}$ are the simulated signal and background yields in the signal region ($M_{\text{bc}} > 5.27 \gevcc$ and ${-0.10 < \Delta E < 0.06 \gev}$) and relax it slightly for subsequent analysis since the BDT output is included in the fit of sample composition.
This requirement rejects $99\%$ ($98\%$) of background and retains $71\%$ ($68\%$) of ${\Bp \to \pip\piz}$ (${\Bp \to \Kp\piz}$) signal. 
A systematic uncertainty on the signal efficiency of the BDT selection is evaluated in Sec.~\ref{sec:syst} to account for data-simulation discrepancies in the distribution of the input variables. 
For the fit, we log-transform the continuum suppression output.
The log-transformation is calculated as follows: 
\begin{equation}
C' = \ln{\left(\frac{C - \text{BDT$_{\text{req}}$}}{1 - C}\right)},
\end{equation}
where $C$ is the continuum suppression output and  $\text{BDT$_{\text{req}}$}$ is the selection requirement placed on the output.

\section{Fit Modeling}

To determine signal yields, a three-dimensional ($\Delta E$, $M_{\text{bc}}$, $C'$) simultaneous extended unbinned maximum likelihood fit to the ${\Bp \to \Kp \piz}$ and ${\Bp \to \pip \piz}$ samples is performed.
To determine the {\it CP} asymmetry, independent signal yields for the negatively and positively charged $B$ candidates are fitted simultaneously.
The fit assumes no correlation between the fit observables.
Along with signal, three background components are considered in the fit: 
continuum-background, background from $\FourS \to \BB$ processes, and feed-across background.
The latter background consists of ${\Bp \to \Kp \piz}$ events that are incorrectly reconstructed as ${\Bp \to \pip \piz}$ or vice versa.

The shapes of the individual fit components are determined empirically from simulation, which is corrected as a function of particle momentum and polar angle to account for particle-identification discrepancies with data.
The signal and feed-across $\Delta E$ distributions are modeled with the sum of a Gaussian function and a function consisting of a Gaussian core with power-law tails at low and high $\Delta E$ values \cite{Oreglia:1980cs}.
The continuum-background $\Delta E$ distribution is modeled with a first (second) order Chebyshev polynomial for ${\Bp \to \Kp \piz}$ (${\Bp \to \pip \piz}$). 
The $\Delta E$ distribution for $\BB$ background is modeled with a kernel density estimator.
The signal and feed-across $M_{\text{bc}}$ distributions are modeled using a function consisting of a Gaussian core with a power-law tail at low $M_{\text{bc}}$ values. 
For the ${\Bp \to \pip \piz}$ feed-across component an additional Gaussian function is added. 
The continuum-background $M_{\text{bc}}$ distribution is modeled with the sum of eight ARGUS functions \cite{ALBRECHT1990278} with fixed endpoints evenly spaced from 5.287 to 5.29$\gevcc$.
The other shape parameter are shared among all ARGUS functions. 

The sum of eight ARGUS functions is chosen because the beam energy $E^*$ was subject to systematic shifts of $\mathcal{O}(1 \mev)$ throughout the data taking, impacting the endpoint of the $M_{\text{bc}}$ distribution.
The contribution of each ARGUS function is fixed to the fraction of events collected at center-of-mass energies corresponding to the respective endpoints. 
The remaining parameters of the $\Delta E$ and $M_{\text{bc}}$ shapes for the continuum-background are determined by the fit.
The $\BB$ $M_{\text{bc}}$ distribution is modeled with a kernel density estimator.
The $C'$ distribution for each component is modeled using a JohnsonSU distribution \cite{Johnson:1949zj}, which contains four parameters: a mean, a width, a skewness parameter, and a parameter setting the size of the tail.

To further account for mismodeling between data and simulation, correction parameters are extracted in data using the control channels ${\Bz \to \Dzb (\to \Kp \pim) \piz}$ (for the signal and feed-across $\Delta E$ and $M_{\text{bc}}$ shapes) and ${\Bp \to \Dzb (\to \Kp \pim \piz) \pip}$ (for the signal and feed-across $C'$ shapes) as well as off-resonance data (for the continuum $C'$ shape). 
The correction parameters shift the means and scale the widths of the fit models irrespective of the $\B$ candidate charge. 
The obtained shift parameters for $M_{\text{bc}}$ and $\Delta E$ are small, below $1\mevcc$ and $1\mev$.
The width scaling parameters are approximately 1.04 and 1.16, respectively.
For $C'$ the shift reaches values up to $-0.6$ in units of the log-transformed continuum suppression output and the scaling parameter reaches values up to 1.07.

\section{Determination of Physics Parameters}

We fit directly for raw branching fractions and {\it CP} asymmetries, instead of fitting the ${\Bp \to \Kp \piz}$ and ${\Bp \to \pip \piz}$ signal and feed-across yields.
The individual yields are computed from the physics parameters using the following equation: 

\begin{equation}
    \label{eqn:master_equ}
    N_{h^+ \piz}^{\mp;~\text{sig, bkg}} = N_{\BpBm}~\epsilon_{h^+ \piz}^{\text{sig, bkg}}~\mathcal{B}^{h^+ \piz}_{\text{raw}}~\frac{1 \pm \mathcal{A}_{\text{raw}}^{h^+ \piz}}{2},
\end{equation}
where $h^+$ is either $\Kp$ or $\pip$; $N_{h^+ \piz}^{\mp;~\text{sig, bkg}}$ is the signal or feed-across yield for negatively or positively charged signal $B$ candidates; $N_{\BpBm}$ is the number of produced $\BpBm$ pairs; $\epsilon_{h^+ \piz}^{\text{sig, bkg}}$ is the signal or feed-across reconstruction efficiency; $\mathcal{B}^{h^+ \piz}_{\text{raw}}$ is the measured raw branching fraction; and~$\mathcal{A}_{\text{raw}}^{h^+ \piz}$ is the measured charge-yield asymmetry, defined as $\mathcal{A}_{\text{raw}}^{h^+ \piz} = \frac{N(\Bm) - N(\Bp)}{N(\Bm) + N(\Bp)}$, with $N(\Bm)$ ($N(\Bp)$) being the number of negatively (positively) charged $B$ mesons.

For the continuum and \BB background, yields are determined by the fit independently for each charge. 

To summarize, 19 parameters are determined by the fit:
the $\mathcal{B}_{\text{raw}}$ values for ${\Bp \to \pip \piz}$ and ${\Bp \to \Kp \piz}$ (2 parameters), the asymmetries $\mathcal{A}_{\text{raw}}$ for ${\Bp \to \pip \piz}$ and ${\Bp \to \Kp \piz}$ (2 parameters), the continuum-background shape parameters for $\Delta E$ and $M_{\text{bc}}$ (7 parameters), the continuum-background yields separate for each sample and charge (4 parameters), and the $\BB$ yields separate for each sample and charge (4 parameters).

\section{Inputs and efficiencies}

The number of $\BpBm$ pairs is calculated from the number of produced $\BB$ pairs (198~million) assuming a ratio of $\FourS \to \BpBm$ and $\FourS \to \BB$ of $0.514 \pm 0.006$ \cite{PhysRevD.98.030001}. The reconstruction efficiency for the ${\Bp \to \pip \piz}$ (${\Bp \to \Kp \piz}$) signal is estimated to be $31.8\%$ ($26.9\%$) using simulation.
The reconstruction efficiency for the respective feed-across background is estimated to be $5.3\%$ ($6.3\%$).

The measured charge-dependent yield asymmetries
$$\mathcal{A}_{\text{raw}}^{h^+ \piz} = \mathcal{A}_{{\it CP}}^{h^+ \piz} +  \mathcal{A}_{\text{det}}^{h^+ \piz},$$
are the sum of the {\it CP} asymmetry $\mathcal{A}_{{\it CP}}^{h^+ \piz}$ and the instrumental asymmetry $\mathcal{A}_{\text{det}}^{h^+ \piz}$ due to differences in interaction and reconstruction probabilities between particles and antiparticles.
We estimate the instrumental asymmetry for charged pions, $\mathcal{A}_{\text{det}}(\pip) = -0.005 \pm 0.010$, by measuring the charge asymmetry in an abundant sample of $\Dp \to \KS \pip$ decays assuming negligible contributions from $\KS$ asymmetries and subtracting the known $\mathcal{A}_{{\it CP}}(\Dp \to \KS \pip)$ value~\cite{PhysRevD.98.030001}.
To obtain the instrumental asymmetry for charged kaons $\mathcal{A}_{\text{det}}(\Kp)$, we determine the charge asymmetry in ${\Dz \to \Km \pip}$ decays, which provides the joint $\Km\pip$ instrumental asymmetry $\mathcal{A}_{\text{det}}(\Km\pip)$.
In $\Dz \to \Km \pip$ decays, direct {\it CP} violation is expected to be smaller than~$0.1\%$~\cite{PhysRevD.98.030001}. We therefore attribute any nonzero asymmetry to instrumental charge asymmetries.
Combining $\mathcal{A}_{\text{det}}(\Km\pip)$ with $\mathcal{A}_{\text{det}}(\pip)$, we obtain the $\Kp$ instrumental asymmetry, $\mathcal{A}_{\text{det}}(\Kp) = -0.011 \pm 0.010$.

\section{Fit results}

\begin{figure}
    \centering
    \includegraphics[width=0.49\textwidth]{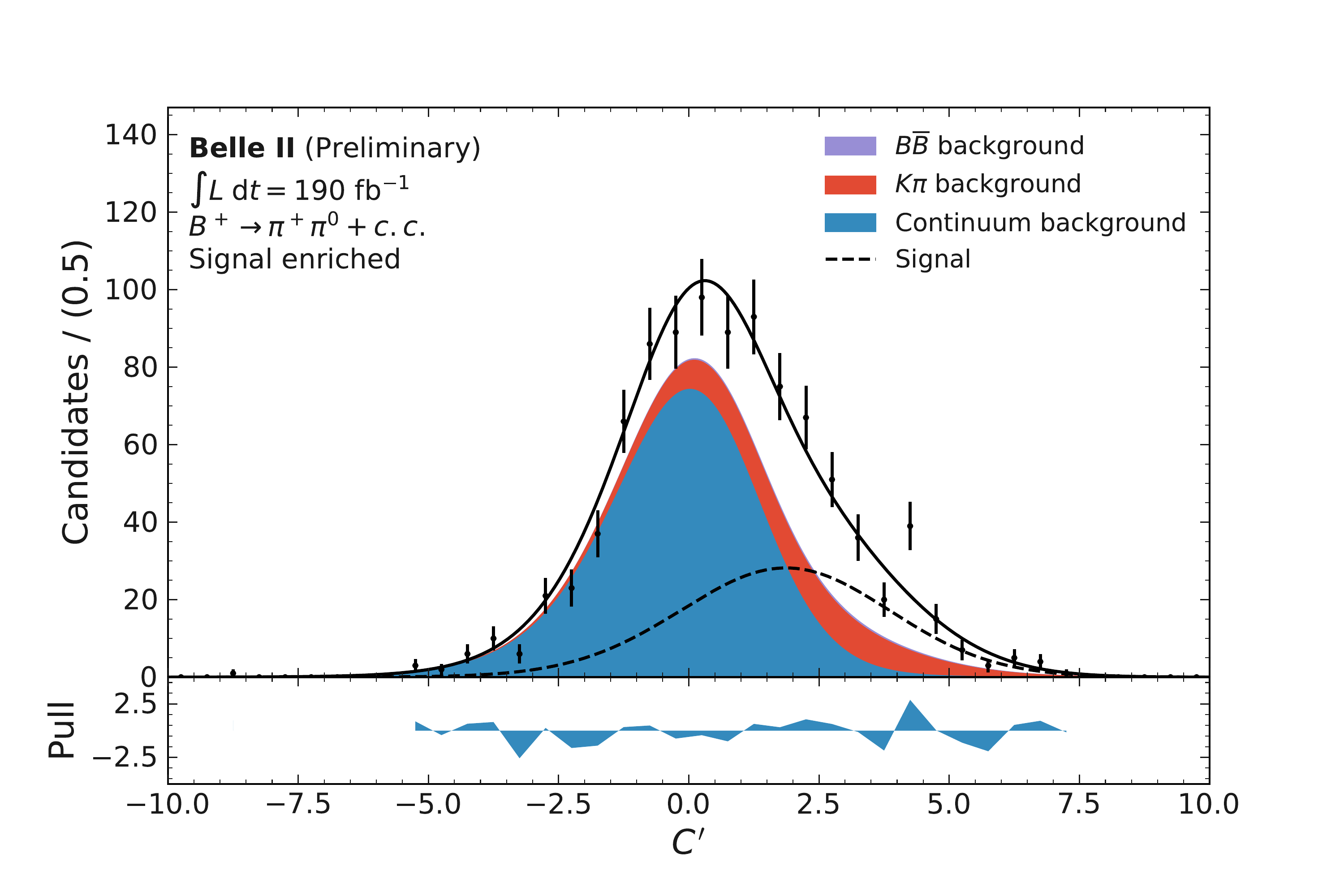}
    \includegraphics[width=0.49\textwidth]{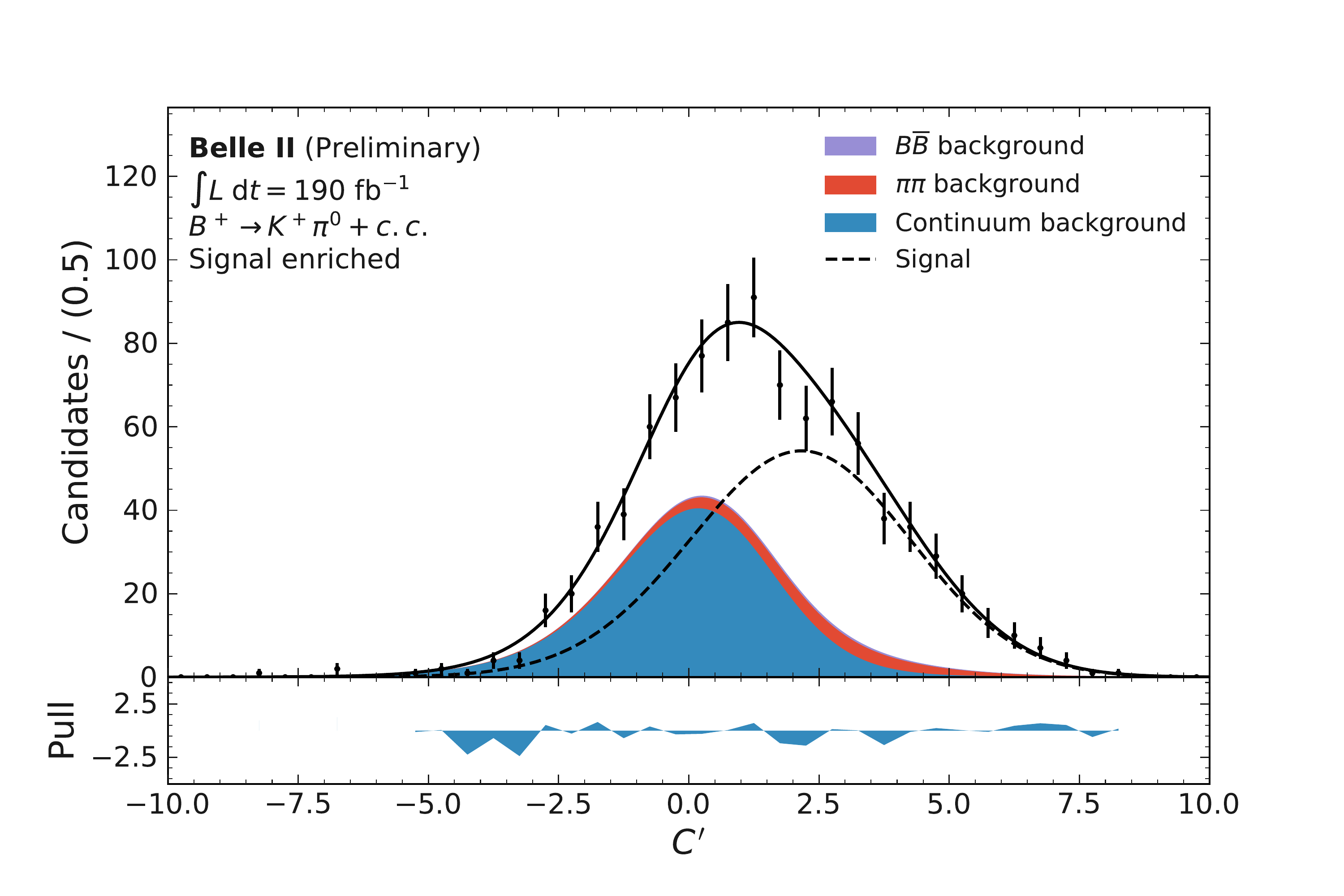}
    \includegraphics[width=0.49\textwidth]{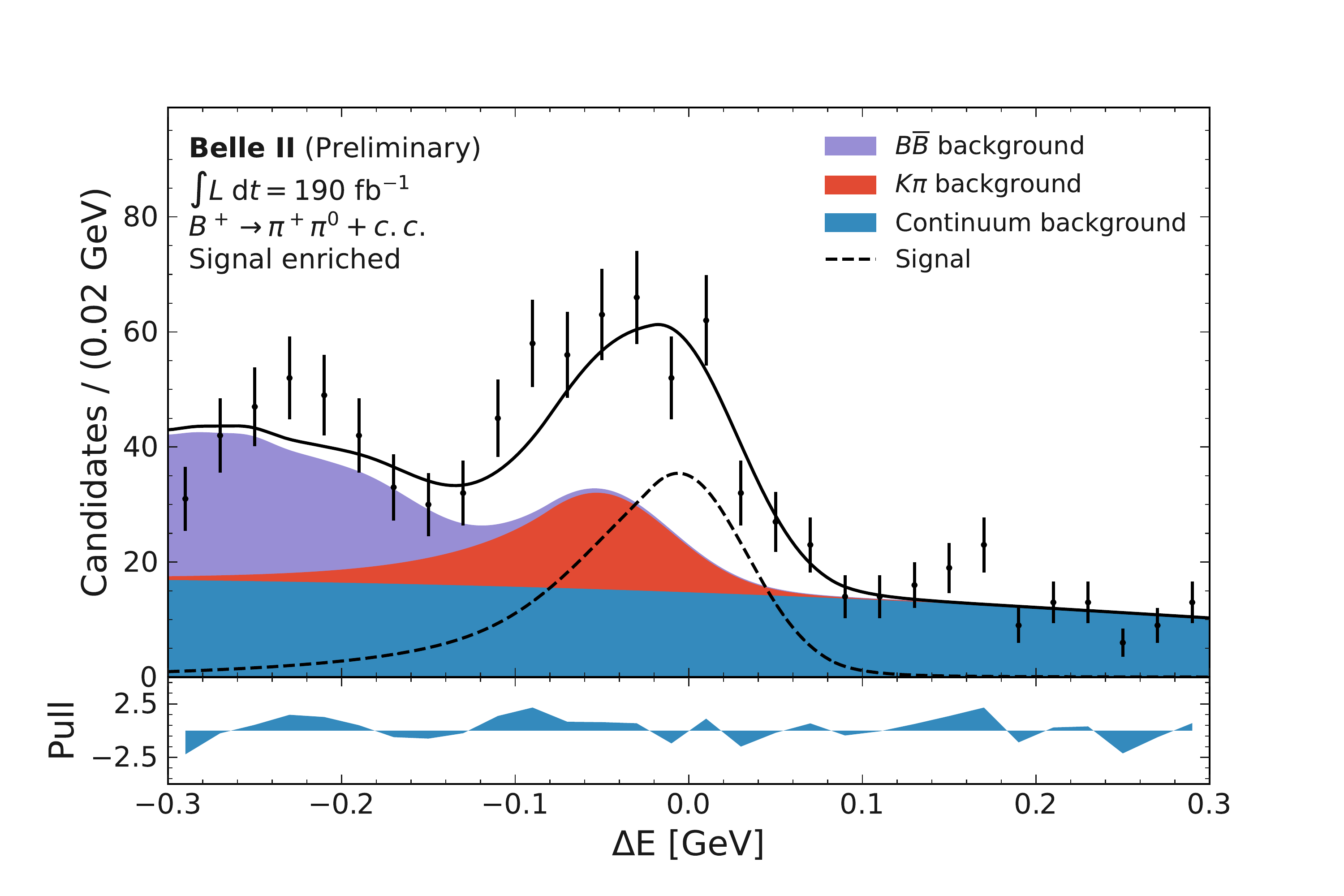}
    \includegraphics[width=0.49\textwidth]{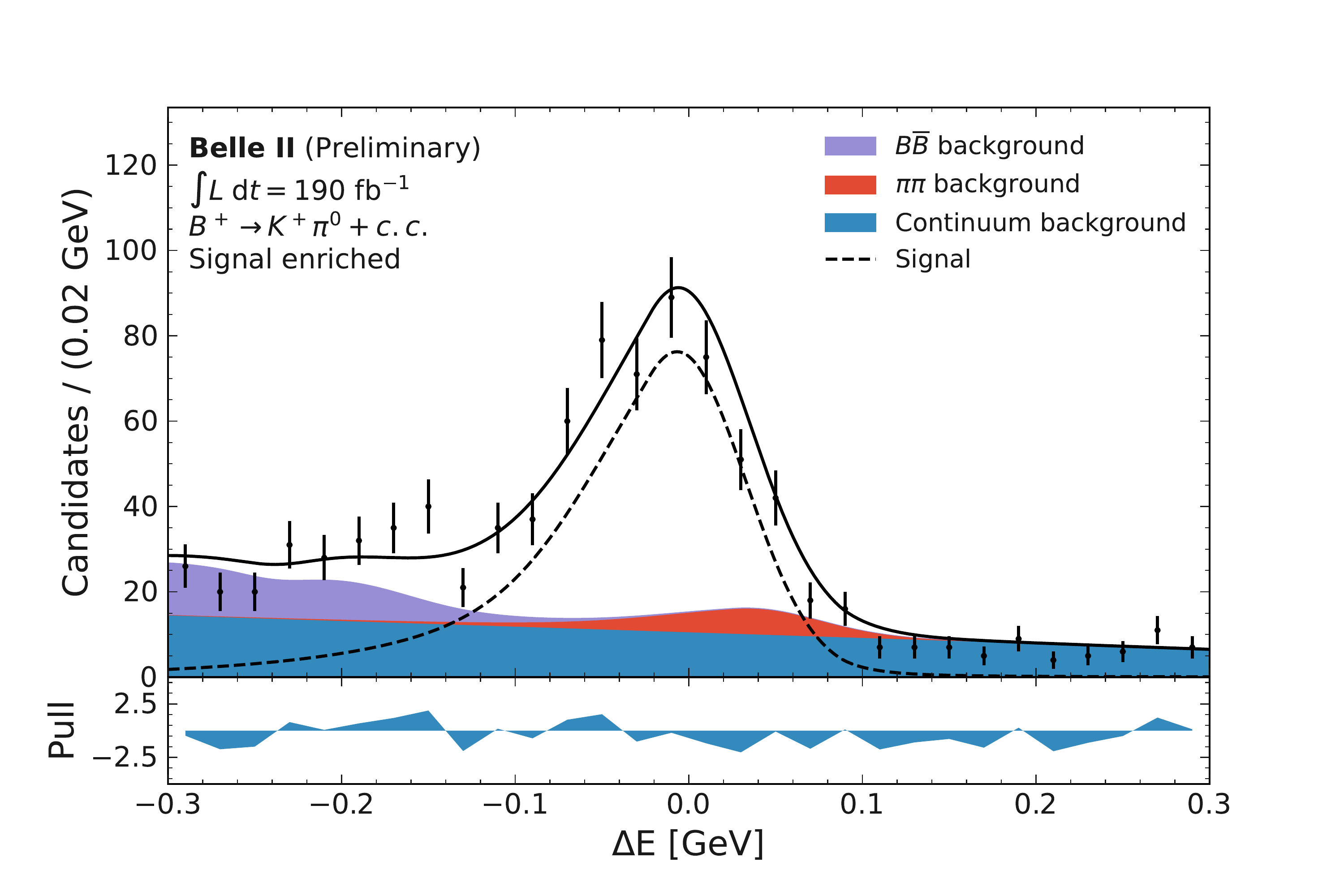}
    \includegraphics[width=0.49\textwidth]{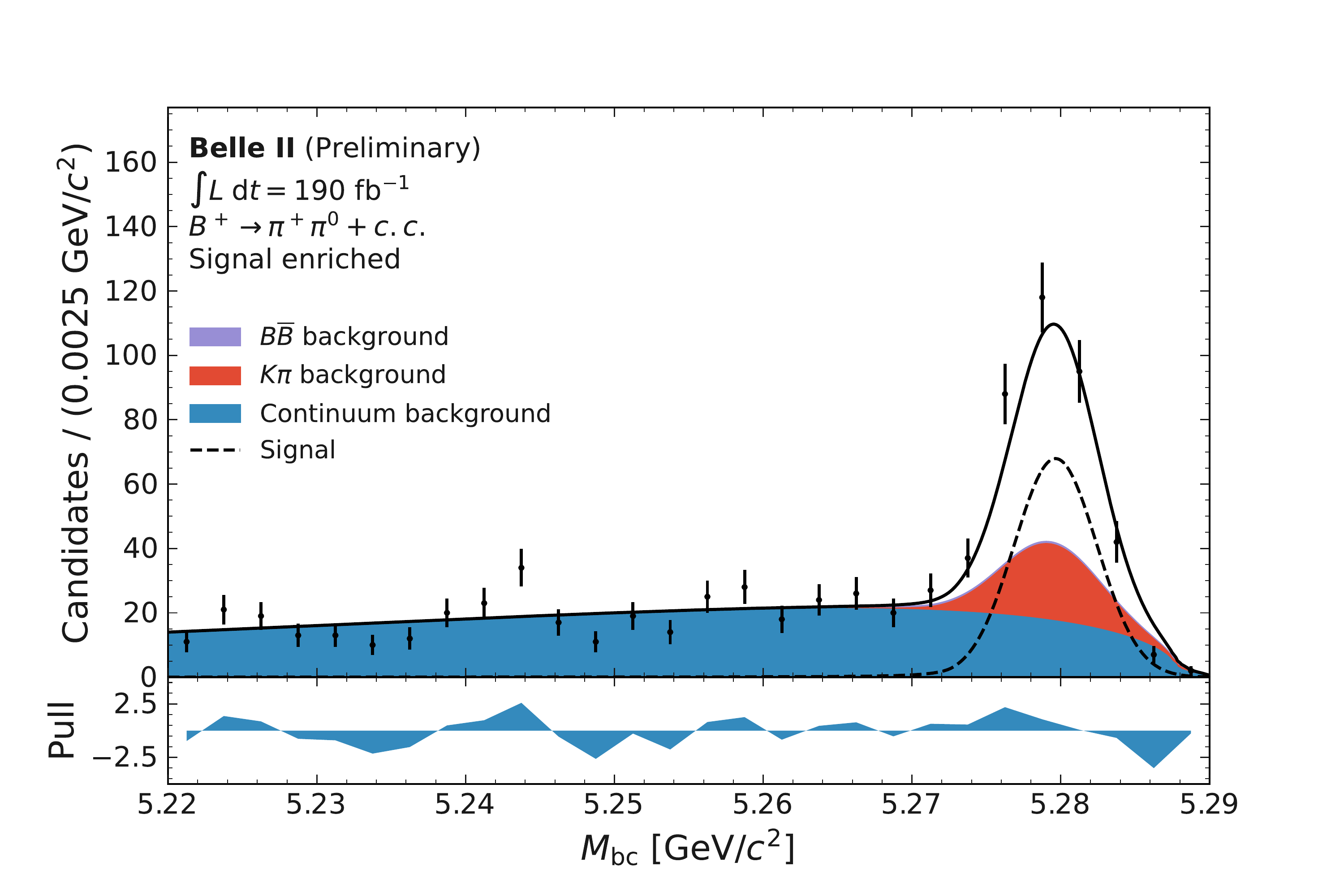}
    \includegraphics[width=0.49\textwidth]{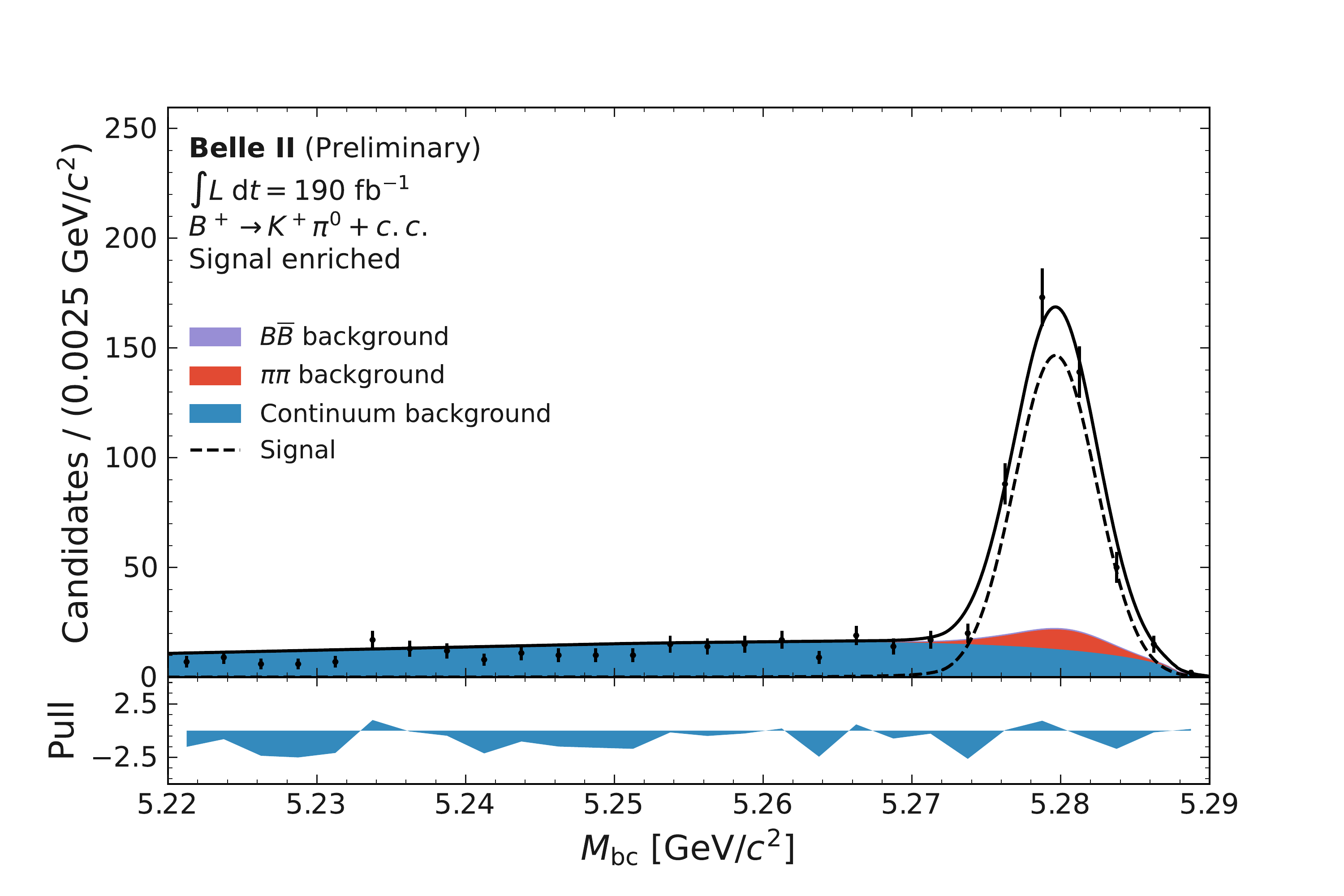}
    \caption{Distribution of the log transformed BDT output $C'$ (top), $\Delta E$ (middle), and $M_{\text{bc}}$ (bottom) of ${\Bp \to \pip \piz}$ (left) and ${\Bp \to \Kp \piz}$ (right) candidates, restricted to signal-enhanced regions selected through ${-0.10 < \Delta E < 0.06\gev}$, $M_{\text{bc}} > 5.27\gevcc$, and $C' > 1$ if the respective variable is not plotted.
    The result of a fit to the sample is shown as a solid black curve. The fit components are shown as black dashed curve (signal), purple shaded area ($\BB$ background), red shaded area (feed-across background), and blue shaded area (continuum-background).
    Differences between observed data and total fit results, normalized by fit uncertainties, (pulls) are also shown.}
    \label{fig:result}
\end{figure}

\begin{figure}
    \centering
    \includegraphics[width=0.49\textwidth]{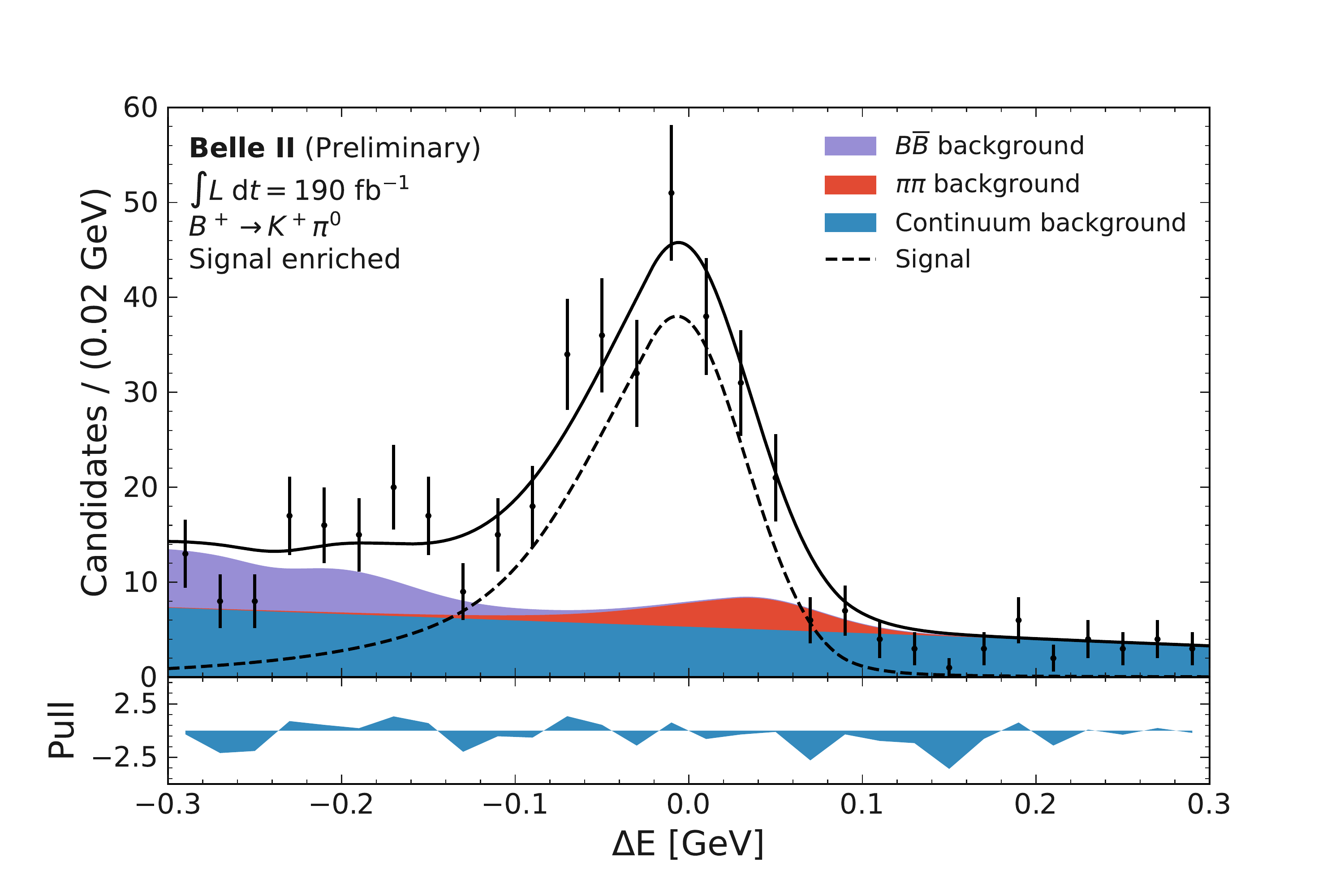}
    \includegraphics[width=0.49\textwidth]{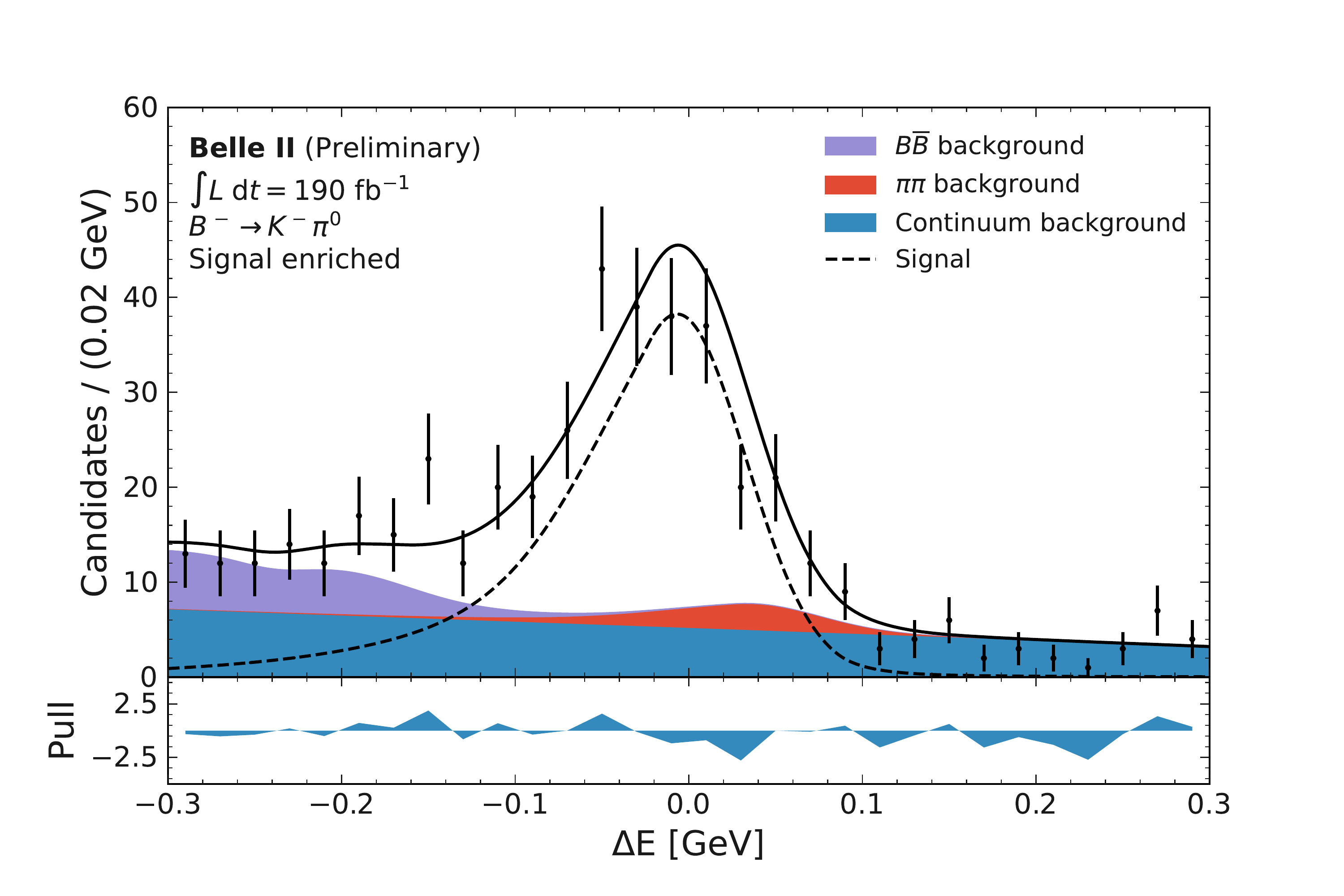}
    \includegraphics[width=0.49\textwidth]{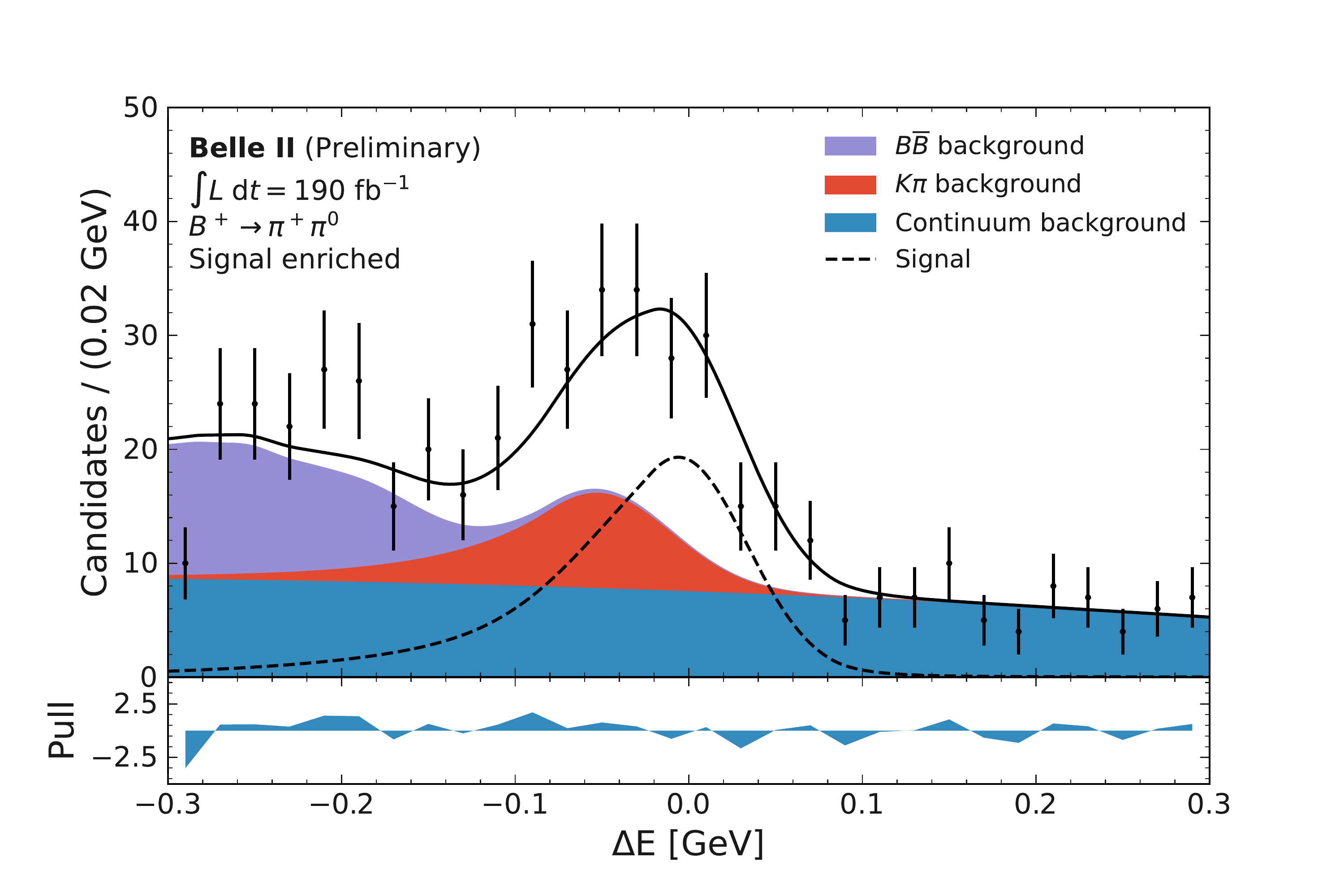}
    \includegraphics[width=0.49\textwidth]{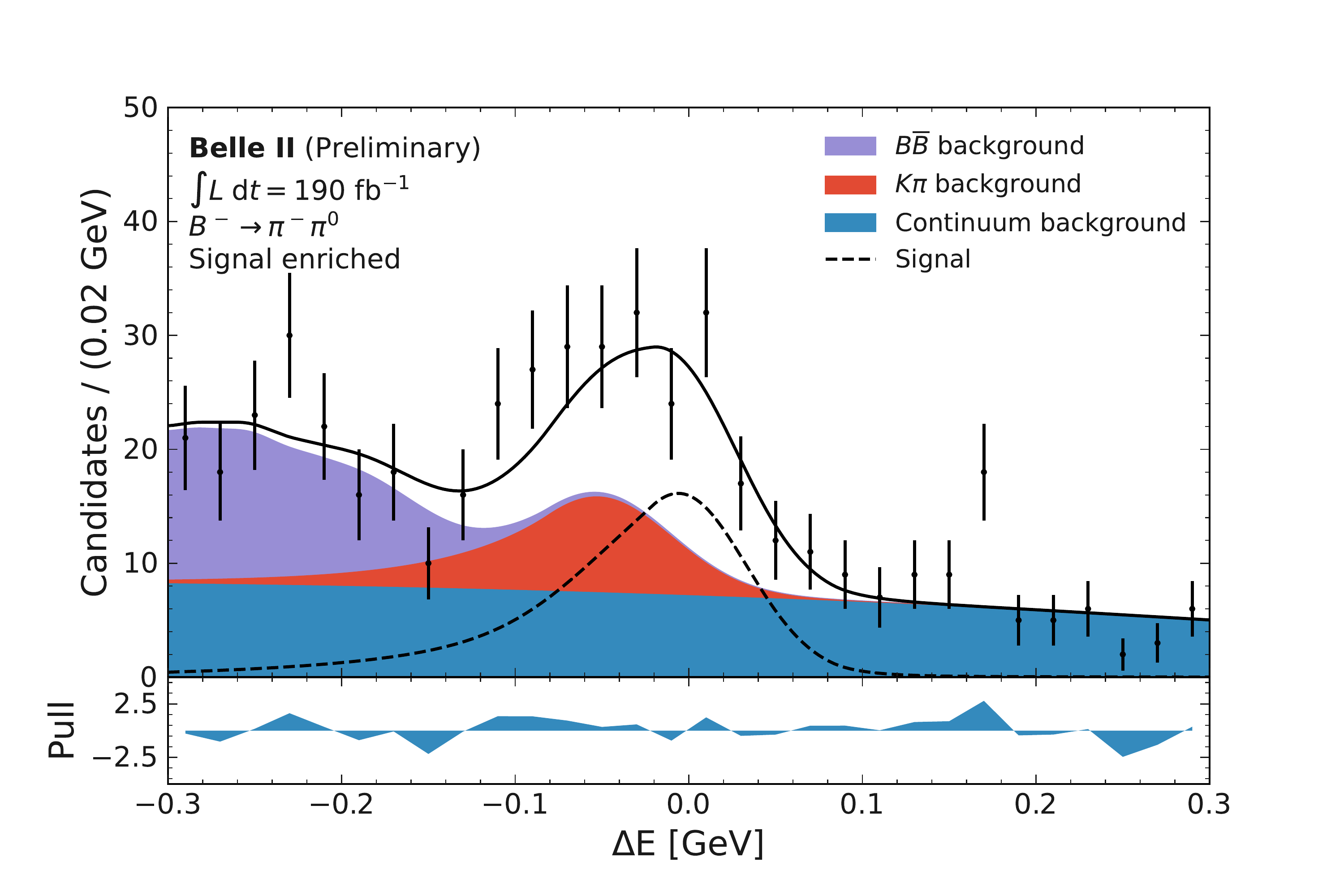}
    \caption{{Distribution of $\Delta E$ for positively charged (left) and negatively charged (right) ${\Bp \to \Kp \piz}$~(top) and ${\Bp \to \pip \piz}$ candidates (bottom), restricted to a signal-enhanced region selected through $M_{\text{bc}} > 5.27\gevcc$ and $C' > 1$.
    The result of a fit to the sample is shown as a solid black curve. The fit components are shown as black dashed curve (signal), purple shaded area ($\BB$~background), red shaded area (feed-across background), and blue shaded area (continuum-background).
    Differences between observed data and total fit results, normalized by fit uncertainties,~(pulls) are also shown.}}
    \label{fig:charge_sep_result}
\end{figure}

In Fig~\ref{fig:result}, charge-integrated signal-enhanced data are shown with fit projections overlaid.
The signal-enhanced region is defined as $M_{\text{bc}} > 5.27 \gevcc$, $-0.1 < \Delta E < 0.06 \gev$, and $C' > 1$.
In addition, charge-specific $\Delta E$ distributions with fit projections overlaid are shown in Fig.~\ref{fig:charge_sep_result} to illustrate the magnitude of $\mathcal{A}_{\text{raw}}^{h^+ \piz}$.
The fit finds raw branching fractions of ${\mathcal{B}_{\text{raw}}(\pip \piz) = (5.58 \pm 0.49)\times 10^{-6}}$ and ${\mathcal{B}_{\text{raw}}(\Kp \piz) = (13.12 \pm 0.63)\times 10^{-6}}$ and asymmetries of ${\mathcal{A}_{\text{raw}}(\pip \piz) = -0.090 \pm 0.085}$ and ${\mathcal{A}_{\text{raw}}(\Kp \piz) = 0.003 \pm 0.047}$, where all uncertainties are statistical.

After correcting for the continuum suppression efficiency (see Sec. \ref{sec:syst}) as well as the instrumental asymmetries, we measure branching fractions of ${\mathcal{B}(\pip \piz) = (6.12 \pm 0.53)\times 10^{-6}}$ and ${\mathcal{B}(\Kp \piz) = (14.30 \pm 0.69)\times 10^{-6}}$ and {\it CP} asymmetries of ${\mathcal{A}_{{\it CP}}(\pip \piz) = -0.085 \pm 0.085}$ and ${\mathcal{A}_{{\it CP}}(\Kp \piz) = 0.014 \pm 0.047}$, where all uncertainties are statistical.
The correlations between physics parameters are summarized in Tab.~\ref{tab:corr_fit}.

\begin{table}[]
    \centering
    \caption{Summary of the correlation between physics parameters.}
    \label{tab:corr_fit}
    \begin{tabular}{c r r r r}
    \hline\hline
         & $\mathcal{B}(\pip \piz)$ & $\mathcal{B}(\Kp \piz)$ & $\mathcal{A}(\pip \piz)$ & $\mathcal{A}(\Kp \piz)$ \\
         \hline
    $\mathcal{B}(\pip \piz)$ & $1.00$ & $-0.32$ & $0.04$ & $0.01$ \\
    $\mathcal{B}(\Kp \piz)$ & & $1.00$ & $-0.02$ & $-0.01$ \\
    $\mathcal{A}(\pip \piz)$ & & & $1.00$ & $-0.33$ \\
    $\mathcal{A}(\Kp \piz)$ & & & & $1.00$ \\
    \hline\hline
    \end{tabular}
\end{table}

\clearpage

\section{Systematic Uncertainties}
\label{sec:syst}

In Table~\ref{tab:syst_br} and \ref{tab:syst_acp} the systematic and statistical uncertainties are summarized. 
The individual sources of systematic uncertainty are discussed in more detail below.

\begin{table}[]
    \caption{Summary of the fractional uncertainties on the branching fractions.}
    \label{tab:syst_br}
    \centering
    \begin{tabular}{l c c}
    \hline
    \hline
    Source & $\Bp \to \Kp \piz$ [\%] & $\Bp \to \pip \piz$ [\%]\\
    \hline
    Tracking & 0.30 & 0.30 \\
    $\B$ counting & 1.5 & 1.5 \\
    $R(\BpBm)$ & 1.2 & 1.2 \\
    $\piz$ efficiency & 4.4 & 4.4 \\
    CS efficiency & 0.9 & 1.1 \\
    Particle identification & 0.2 & 0.5\\
    Multiple candidates & 0.01 & 0.9\\
    Continuum BDT shift and scale (\Kp \piz) & 0.5 & 0.08 \\
    Continuum BDT shift and scale (\pip \piz) & 0.1 & 1.6 \\
    $\Delta E$ shift and scale & 2.0 & 6.3 \\
    $M_{\text{bc}}$ shift and scale & 1.1 & 2.3 \\
    Signal BDT shift and scale (\Kp \piz) & 0.4 & 0.1 \\
    Signal BDT shift and scale (\pip \piz) & 0.02 & 0.8 \\
    $\BB$ Shape & 0.4 & 0.2 \\
    \hline
    Total systematic uncertainty & 5.5 & 8.6 \\
    \hline
    Statistical uncertainty & 4.8  & 8.7 \\
    \hline
    \hline
    \end{tabular}
\end{table}

\begin{table}[]
    \caption{Summary of the absolute uncertainties on the {\it CP} asymmetries.}
    \label{tab:syst_acp}
    \centering
    \begin{tabular}{l c c}
    \hline
    \hline
    Source & $\Bp \to \Kp \piz$ & $\Bp \to \pip \piz$ \\
    \hline
    Continuum BDT shift and scale (\Kp \piz) & 0.0002 & 0.0006 \\
    Continuum BDT shift and scale (\pip \piz) & 0.0010 & 0.0092 \\
    $\Delta E$ shift and scale & 0.0014 & 0.0038 \\
    $M_{\text{bc}}$ shift and scale & 0.0008 & 0.0023 \\
    Signal BDT shift and scale (\Kp \piz) & 0.0002 & $< 0.0001$ \\
    Signal BDT shift and scale (\pip \piz) & 0.0002 & 0.0005 \\
    $\BB$ Shape & 0.0000 & 0.0001 \\
    Instrumental asymmetry & 0.010 & 0.010 \\
    Fit bias & - & 0.0118\\
    \hline
    Total systematic uncertainty & 0.0102 & 0.0185\\
    \hline
    Statistical uncertainty & 0.0470 & 0.0851 \\
    \hline
    \hline
    \end{tabular}
\end{table}

We assess a systematic uncertainty associated with possible data-simulation discrepancies in the reconstruction of charged particles. The tracking efficiency in data agrees with the value observed in simulation within a 0.30\% uncertainty, which is assumed as uncertainty on the branching fraction results.
We assign a 1.5\% systematic uncertainty on the number of $\BB$ pairs, which includes the selection efficiency, integrated luminosity, and potential shifts from the peak center-of-mass energy during data taking.
We assign the uncertainty of the ratio $R(\BpBm)$ of $\FourS \to \BpBm$ and $\FourS \to \BB$ as a systematic uncertainty \cite{PhysRevD.98.030001}.

Several systematic uncertainties are related to the selection and reconstruction efficiencies. 
The $\piz$ reconstruction efficiency is assessed using the decays $\Dz \to \Kp \pim \piz$ and ${\Dz \to \Kp \pim}$. 
We compare the yields from fits to the $\Dz$ invariant mass distribution and obtain $0.950 \pm 0.042$ for the ratio of the $\piz$ reconstruction efficiency in simulation and in data. 
This ratio is compatible with unity. The uncertainty on this ratio is used as a systematic uncertainty.
The efficiency of the continuum-suppression requirement is studied using the control channel ${\Bp \to \Dzb (\to \Kp \pim \piz) \pip}$.
The ratio of the efficiency found in data divided by the efficiency found in the simulation is $0.9124 \pm 0.0099$ ($0.9178 \pm 0.0082$) for the $\Bp \to \pip \piz$ ($\Bp \to \Kp \piz$) BDT.
The ratios are incompatible with unity. 
We scale the branching fractions by these ratios and assign the uncertainty on the ratios as systematic uncertainties. 
The particle-identification corrections are varied 100 times within their uncertainty and used to derive alternative reconstruction efficiencies.
The fit to data is repeated with the alternative reconstruction efficiencies. 
The standard deviation of the distribution of the physics parameter is assigned as a systematic uncertainty.
In 1\% of events in which multiple candidates are reconstructed, we retain all candidates. 
To assess a systematic uncertainty associated with a possible data-simulation mismatch in the efficiency of this criterion, we repeat the fit to the data by randomly selecting a single candidate in each event.
The difference from the nominal fit result is taken as a systematic uncertainty.

Systematic uncertainties associated with the correction factors (\emph{i.e.}, the shift and scaling parameters) are assessed by repeating the fit on the experimental data 100 times with alternative correction parameters.
For each of the 100 fits, the correction factors are drawn within their uncertainties from a Gaussian distribution, taking correlations into account.
Each fit yields a value of the physics parameters (\textit{e.g.}, $\mathcal{A}_{{\it CP}}$) slightly different from that resulting from the nominal fit. 
The standard deviation of the $\mathcal{A}_{{\it CP}}$ distribution is assigned as the systematic uncertainty for the correction factors.
This procedure is repeated for all correction factors, namely those of the continuum BDT shapes, the $\Delta E$ shapes, the $M_{\text{bc}}$ shapes, and the signal BDT shapes. 
To assess a systematic uncertainty for the $\BB$ shape, we develop an alternative fit model for the $\BB$ background. 
We generate 100 simplified simulated data sets around this alternative fit model and fit the data sets with the alternative and nominal fit model. 
The mean of the differences between the fit results is assigned as systematic uncertainty. 

We consider the uncertainty on the instrumental asymmetries $\mathcal{A}_{\text{det}}$ as a systematic uncertainty. 
The instrumental asymmetries depend on the kinematic properties of the relevant charged particles and on the number of associated hits. The value of $\mathcal{A}_{\text{det}}$ is obtained from control modes in which tracks are selected to have kinematic and hit-multiplicity distributions as close as possible to those of the signal. The uncertainty on $\mathcal{A}_{\text{det}}$ is dominated by  possible residual differences. 

To validate the fit of sample composition, we perform simplified simulated experiments 
in which the true values of either $\mathcal{B}$ or $\mathcal{A}_{{\it CP}}$ are varied in 20\% increments from $60\%$ to $140\%$.
This validation shows a small bias for the {\it CP} asymmetry of ${\Bp \to \pip \piz}$. 
We assign this bias as systematic uncertainty.

\section{Summary}

We report a measurement of the branching fractions and {\it CP} asymmetries of ${\Bp \to \pip \piz}$ and ${\Bp \to \Kp \piz}$ decays.
The results are based on data corresponding to an integrated luminosity of $190\invfb$ recorded by the Belle~II detector at the $\FourS$ resonance.
We measure 
\begin{center}
${\mathcal{B}(\Bp \to \pip \piz) = (6.12 \pm 0.53 \pm 0.53)\times 10^{-6}}$,

${\mathcal{B}(\Bp \to \Kp \piz) = (14.30 \pm 0.69 \pm 0.79)\times 10^{-6}}$,

${\mathcal{A}_{{\it CP}}(\Bp \to \pip \piz) = -0.085 \pm 0.085 \pm 0.019}$,  

${\mathcal{A}_{{\it CP}}(\Bp \to \Kp \piz) = 0.014 \pm 0.047 \pm 0.010}$, 
\end{center}
where the first uncertainties are statistical and the second are systematic.
The results improve and supersede a previous Belle~II measurement \cite{Belle-II:2021gkm} and agree with current world averages \cite{PhysRevD.98.030001}.

\bibliography{belle2}

\providecommand{\href}[2]{#2}\begingroup\raggedright\begin{thebibliography}{10}

\bibitem{CC}
{{Charge conjugation is implied throughout this note unless otherwise stated}}.

\bibitem{Charles_2017}
J.~Charles, O.~Deschamps, S.~Descotes-Genon, and V.~Niess, {\em Isospin
  analysis of charmless B-meson decays\/},
  \href{http://dx.doi.org/10.1140/epjc/s10052-017-5126-9}{The European Physical
  Journal C {\bf 77} (2017) no.~8, }.

\bibitem{PhysRevD.59.113002}
M.~Gronau and J.~L. Rosner, {\em Combining $\mathrm{CP}$ asymmetries in $B \to
  K\pi$ decays\/},  \href{http://dx.doi.org/10.1103/PhysRevD.59.113002}{Phys.
  Rev. D {\bf 59} (1999)  113002}.
  \url{https://link.aps.org/doi/10.1103/PhysRevD.59.113002}.

\bibitem{Aaij_2021}
R.~Aaij et al., {\em Measurement of CP violation in the decay $\Bp \to \Kp
  \piz$\/},  \href{http://dx.doi.org/10.1103/physrevlett.126.091802}{Physical
  Review Letters {\bf 126} (2021) no.~9, }.

\bibitem{HFLAV}
Y.~Amhis et al., {\em Averages of $b$-hadron, $c$-hadron, and $\tau$-lepton
  properties as of 2021\/},  \href{http://arxiv.org/abs/2206.07501}{{\tt
  arXiv:2206.07501}}.

\bibitem{Gershon_2007}
T.~Gershon and A.~Soni, {\em Null tests of the Standard Model at an
  International Super $B$ Factory\/},
  \href{http://dx.doi.org/10.1088/0954-3899/34/3/006}{Journal of Physics G:
  Nuclear and Particle Physics {\bf 34} (2007) no.~3, 479--492}.

\bibitem{Belle-II:2021gkm}
F.~Abudin\'en et al., {Belle-II}, {\em {Measurements of branching fractions and
  direct CP-violating asymmetries in $B^+ \to K^+ \pi^0$ and $\pi^+ \pi^0$
  decays using 2019 and 2020 Belle II data}\/},
  \href{http://arxiv.org/abs/2105.04111}{{\tt arXiv:2105.04111 [hep-ex]}}.

\bibitem{Belle-II:2010dht}
T.~Abe et al., {Belle-II}, {\em {Belle II Technical Design Report}\/},
  \href{http://arxiv.org/abs/1011.0352}{{\tt arXiv:1011.0352
  [physics.ins-det]}}.

\bibitem{Kuhr:2018lps}
T.~Kuhr, C.~Pulvermacher, M.~Ritter, T.~Hauth, and N.~Braun, {Belle-II
  Framework Software Group}, {\em {The Belle II Core Software}\/},
  \href{http://dx.doi.org/10.1007/s41781-018-0017-9}{Comput. Softw. Big Sci.
  {\bf 3} (2019) no.~1, 1}.

\bibitem{Jadach:1999vf}
S.~Jadach, B.~F.~L. Ward, and Z.~Was, {\em {The Precision Monte Carlo Event
  Generator $\mathcal{KK}$ For Two-Fermion Final States In $e^+ e^-$
  Collisions}\/},
  \href{http://dx.doi.org/10.1016/S0010-4655(00)00048-5}{Comput. Phys. Commun.
  {\bf 130} (2000)  260--325}.

\bibitem{Sjostrand:2014zea}
T.~Sj\"ostrand, S.~Ask, J.~R. Christiansen, R.~Corke, N.~Desai, P.~Ilten,
  S.~Mrenna, S.~Prestel, C.~O. RasmPut a reference to flavor tagging~paper
  ussen, and P.~Z. Skands, {\em {An introduction to PYTHIA 8.2}\/},
  \href{http://dx.doi.org/10.1016/j.cpc.2015.01.024}{Comput. Phys. Commun. {\bf
  191} (2015)  159--177}, \href{http://arxiv.org/abs/1410.3012}{{\tt
  arXiv:1410.3012 [hep-ph]}}.

\bibitem{Lange:2001uf}
D.~J. Lange, {\em {The EvtGen particle decay simulation package}\/},
  \href{http://dx.doi.org/10.1016/S0168-9002(01)00089-4}{Nucl. Instrum. Meth. A
  {\bf 462} (2001)  152--155}.

\bibitem{GEANT4:2002zbu}
S.~Agostinelli et al., {GEANT4}, {\em {GEANT4--a simulation toolkit}\/},
  \href{http://dx.doi.org/10.1016/S0168-9002(03)01368-8}{Nucl. Instrum. Meth. A
  {\bf 506} (2003)  250--303}.

\bibitem{Lewis:2018ayu}
P.~M. Lewis et al., {\em {First Measurements of Beam Backgrounds at
  SuperKEKB}\/},  \href{http://dx.doi.org/10.1016/j.nima.2018.05.071}{Nucl.
  Instrum. Meth. A {\bf 914} (2019)  69--144}.

\bibitem{COM}
{{Unless otherwise stated, quantities denoted by a star symbol (*) are
  estimated in the center-of-mass frame}}.

\bibitem{modMbc}
{{This definition of $M_{\text{bc}}$ differs from the definition usually used
  at Belle~II. This is due to the high correlation between $M_{\text{bc}}$ and
  $\Delta E$ for two body final states with a $\piz$. The correlations can be
  reduced by using this modified definition of $M_{\text{bc}}$
  \cite{PhysRevD.87.031103}}}.

\bibitem{fw}
G.~C. Fox and S.~Wolfram, {\em Observables for the Analysis of Event Shapes in
  ${e}^{+}{e}^{\ensuremath{-}}$ Annihilation and Other Processes\/},
  \href{http://dx.doi.org/10.1103/PhysRevLett.41.1581}{Phys. Rev. Lett. {\bf
  41} (1978)  1581--1585}.

\bibitem{CLEO:1995rok}
D.~M. Asner et al., {CLEO}, {\em {Search for exclusive charmless hadronic B
  decays}\/},  \href{http://dx.doi.org/10.1103/PhysRevD.53.1039}{Phys. Rev. D
  {\bf 53} (1996)  1039--1050}.

\bibitem{Belle:ksfw}
S.~H. Lee et al., {Belle}, {\em {Evidence for $B^0 \to \pi^0 \pi^0$}\/},
  \href{http://dx.doi.org/10.1103/PhysRevLett.91.261801}{Phys. Rev. Lett. {\bf
  91} (2003)  261801}.

\bibitem{Belle-II:2021zvj}
F.~Abudin\'en et al., {Belle-II}, {\em {$B$-flavor tagging at Belle II}\/},
  \href{http://dx.doi.org/10.1140/epjc/s10052-022-10180-9}{Eur. Phys. J. C {\bf
  82} (2022) no.~4, 283}, \href{http://arxiv.org/abs/2110.00790}{{\tt
  arXiv:2110.00790 [hep-ex]}}.

\bibitem{Oreglia:1980cs}
M.~Oreglia et al., {\em Study of the reaction
  ${\ensuremath{\psi}}^{\ensuremath{'}}\ensuremath{\rightarrow}\ensuremath{\gamma}\ensuremath{\gamma}\frac{J}{\ensuremath{\psi}}$\/},
  \href{http://dx.doi.org/10.1103/PhysRevD.25.2259}{Phys. Rev. D {\bf 25}
  (1982)  2259--2277}.

\bibitem{ALBRECHT1990278}
H.~Albrecht et al., {\em Search for hadronic b→u decays\/},
  \href{http://dx.doi.org/https://doi.org/10.1016/0370-2693(90)91293-K}{Physics
  Letters B {\bf 241} (1990) no.~2, 278--282}.

\bibitem{Johnson:1949zj}
N.~L. Johnson, {\em {Systems of frequency curves generated by methods of
  translation}\/},
  \href{http://dx.doi.org/10.1093/biomet/36.1-2.149}{Biometrika {\bf 36} (1949)
   149--176}.

\bibitem{PhysRevD.98.030001}
M.~Tanabashi et al., {Particle Data Group}, {\em Review of Particle Physics\/},
   \href{http://dx.doi.org/10.1103/PhysRevD.98.030001}{Phys. Rev. D {\bf 98}
  (2018)  030001}.

\bibitem{PhysRevD.87.031103}
Y.-T. Duh et al., {The Belle Collaboration}, {\em Measurements of branching
  fractions and direct $CP$ asymmetries for
  $B\ensuremath{\rightarrow}K\ensuremath{\pi}$,
  $B\ensuremath{\rightarrow}\ensuremath{\pi}\ensuremath{\pi}$ and
  $B\ensuremath{\rightarrow}KK$ decays\/},
  \href{http://dx.doi.org/10.1103/PhysRevD.87.031103}{Phys. Rev. D {\bf 87}
  (2013)  031103}.

\end{thebibliography}\endgroup
\bibliographystyle{belle2-note}
\end{document}